\input{aipcheck}

\documentclass[
    final            
  ]
  {aipproc}

\layoutstyle{6x9}
\newcommand\msun{\rm{M}_\odot}
\usepackage[]{natbib}

\def\aj{AJ}

\def\apj{ApJ}

\def\apjs{ApJS}

\def\aap{A\&A}

\def\pasp{{PASP}}

\begin{document}

\title{Our Nearest 15 Million Neighbors: The Field Low--Mass Stellar Luminosity and Mass Functions}


\author{John J. Bochanski}{
  address={MIT Kavli Institute for Astrophysics and Space Research, 
77 Massachusetts Avenue,
Cambridge, MA 02139}
, altaddress = {Astronomy Department, University of Washington, Box 351580, U.W.
    Seattle, WA 98195-1580}
}

\author{Suzanne L. Hawley}{
  address={Astronomy Department, University of Washington, Box 351580, U.W.
    Seattle, WA 98195-1580}
}

\author{I. Neill Reid}{
  address={Space Telescope Science Institute, 3700 San Martin Drive
Baltimore, MD 21218}
}

\author{Kevin R. Covey}{
  address={Harvard-Smithsonian Center for Astrophysics,  60 Garden St,  Cambridge, MA 02138 }
}

\author{Andrew A. West}{
  address={MIT Kavli Institute for Astrophysics and Space Research, 
77 Massachusetts Avenue,
Cambridge, MA 02139}
}

\author{David A. Golimowski}{
  address={Space Telescope Science Institute, 3700 San Martin Drive
Baltimore, MD 21218} 
}

\author{\v{Z}eljko Ivezi\'{c}}{
  address={Astronomy Department, University of Washington, Box 351580, U.W.
    Seattle, WA 98195-1580}
}

\begin{abstract}
We report on a new measurement of the luminosity function (LF) and mass function (MF) of field low--mass dwarfs using Sloan Digital Sky Survey (SDSS) photometry.  The final catalog is composed of $\sim$ 15 million low--mass stars (0.1 M$_{\odot} <$ M $<$ 0.8 M$_{\odot}$), spread over 8,400 square degrees.   Distances to the stars are estimated using new photometric parallax relations, constructed from $ugriz$ photometry of nearby low--mass stars with trigonometric parallaxes.  The LF is measured with a novel technique, which simultaneously measures Galactic structure and the stellar LF.  The resulting LF is compared to previous studies and converted to a MF.  The MF is well--described by a log--normal distribution, with M$_{\circ}$ = 0.27 M$_{\odot}$.

\end{abstract}

\maketitle


\section{Introduction}
Low--mass M dwarfs dominate the stellar population of the Milky Way, by number.   These long-lived \cite{1997ApJ...482..420L} and ubiquitous stars comprise $\sim 70\%$ of all stars, yet their small intrinsic luminosities ($L < 0.05$ $L_{\odot}$) have traditionally prohibited the study of large numbers of them.  Fortunately, in recent years, the development of large format CCDs has led to precise photometric surveys of wide solid angles on the sky.  Of note are the Two--Micron All Sky Survey \citep[2MASS; ][]{2006AJ....131.1163S} and the Sloan Digital Sky Survey \citep[SDSS; ][]{2000AJ....120.1579Y}.  These surveys are characterized by large solid angles (thousands of square degrees) coupled with precise ($< 5\%$) and deep ($r \sim 22, $ $J \sim 16.5$) photometry.   The resulting photometric datasets contain millions of low--mass stars, enabling novel investigations employing an unprecedented number of observations. 

Despite the advances made in other cool star topics, two fundamental properties, the luminosity and mass functions, remain uncertain.  The luminosity function (LF), which is directly observable, describes the number density of stars as a function of absolute magnitude (italic $M$, $\Phi(M) = dN / dM$).  The mass function (MF) is typically inferred from the LF, and is defined by the number density in terms of mass (roman M, $\psi({\rm M}) = dN / d{\rm M}$).   For low--mass stars, with lifetimes much greater than the Hubble time, the observed present-day mass function (PDMF) in the field is the initial mass function (IMF).  The uncertainty in these properties can be attributed to fundamental differences among the techniques employed to measure the LF and MF.  Previous investigations have fallen in one of two categories:  nearby, volume--limited studies of trigonometric parallax stars \citep{1997AJ....113.2246R}, or pencil--beam surveys of distant stars over a small solid angle \citep{2001ApJ...555..393Z}.  Sample sizes were limited to a few thousand stars, prohibiting a detailed determination of the IMF.  In general, the IMF has been characterized by a power--law $\psi({\rm M}) = dN / d{\rm M} \propto {\rm M}^{- \alpha}$, \citep{1955ApJ...121..161S} with the exponent $\alpha$ varying over a wide range from 0.5 -- 2.5.  Currently, the largest sample of field stars used to measure the LF and MF of low-mass stars is the study of Covey et al., \citep{covey08}, which studied $\sim 30,000$ stars over 30 sq. degrees.

We present a new measurement of the low-mass stellar LF and MF using SDSS photometry of over 15 million stars.  The observations, calibration and analysis are described below, followed by our results.

\section{Observations and Analysis}
We queried the SDSS catalog archive server (CAS) through the casjobs website\footnote{\url{http://casjobs.sdss.org/CasJobs/}} for point sources in the DR6 - Legacy footprint with the following criteria:
\begin{itemize} 
 \item
 The photometric objects were flagged as \textsc{PRIMARY} with good photometry.  This ensures that objects imaged multiple times were only counted once and the photometric measurement was reliable.
 \item
 The object was classified morphologically as a star.
 \item
 The photometric objects fell within the following brightness and color limits:\hspace{3pt}
\begin{center}
$ i < 22.0, z < 21.2, $ 
$r - i \geq 0.3, i -z \geq 0.2$\\
\end{center}
\end{itemize}
The absolute magnitude of each star in the sample was estimated from new photometric parallax relations derived from $ugriz$ photometry of nearby trigonometric parallax stars \citep{golimowski08}.  

\begin{figure}[!htbp]
\centering
\includegraphics[scale=0.35]{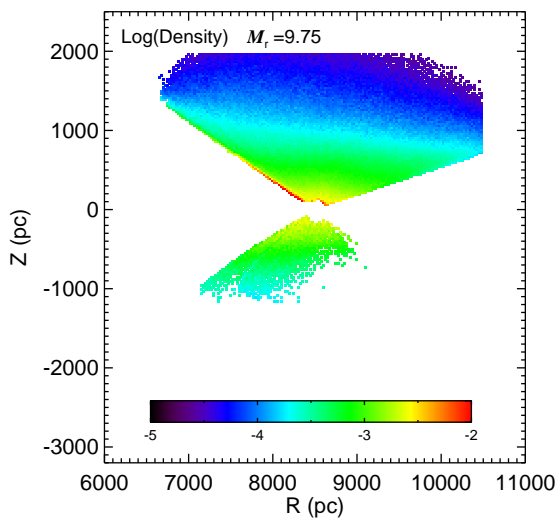} 
\caption{Density (in stars pc$^{-3}$) as a function of Galactic $R$ and $Z$ for a 0.5 mag slice centered on $M_r = 9.75$.  The logarithmic scale is shown beneath the density map.  The disk structure of the Milky Way is clearly evident, with a smooth decline towards larger $R$, and an increase in density approaching the Plane ($Z$ = 0).}
\label{chap5:fig:density_example}
\end{figure}

Most stars in the sample were distant ( $ > 100$ pc),  but traditional pencil--beam survey techniques would not be applicable to a dataset spanning 8,400 sq. degrees.  Thus, we implemented the following method: Absolute magnitudes and distances were computed for each star in our sample.  A small (0.5 mag) slice in absolute magnitude was selected, and the stars were counted as a function of Galactic radius ($R$) and height from the Plane ($Z$).  The volume spanned by SDSS was recorded for each $R, Z$ bin.  The stellar density profile was computed (see Figure \ref{chap5:fig:density_example}), a Galactic model was fit, and the local density of stars was recorded.  This process was repeated for other slices in absolute magnitude, and the LF was constructed from the local density of each slice.  This method computes both the stellar LF and local Galactic structure simultaneously, rather than assuming a Galactic density profile.

\section{Results}
\begin{figure}[!htbp]
\centering
\includegraphics[scale=0.35]{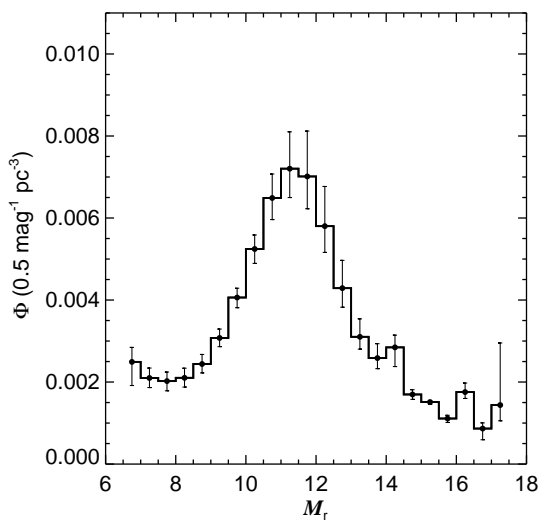} 
\caption{The $M_r$ LF from $M_r=6.75$ to $M_r = 17.25$.  Note the smooth rise to $M_r \sim 11$, with a decline thereafter.}
\label{lf_plot}
\end{figure}

\textbf{Luminosity Function}
Next, we corrected the resulting $M_r$ LF for Malmquist bias and brightening due to unresolved binarity. The resulting LF is shown in Figure \ref{lf_plot}.  The LF peaks near $M_r \sim 11$, corresponding to a spectral type of M3.  The uncertainties in each magnitude bin were estimated by repeating the analysis described above with a full range of photometric parallax relations, which dominate the systematic error budget.

\begin{figure}[!htbp]
\centering
\includegraphics[scale=0.35]{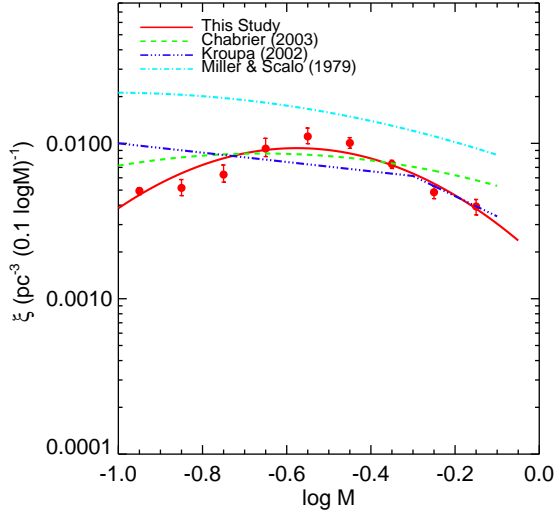} 
\caption{The stellar field MF, from 0.1 $\msun < \rm{M} <$ 0.8$\msun$.  The red points and log-normal fit are the measured data points and best fit from this study.  Canonical IMFs \citep{2003PASP..115..763C,2002Sci...295...82K,1979ApJS...41..513M} are overplotted for comparison, with their colors given in the legend. }
\label{mf_plot}
\end{figure}

\hspace{-20pt}
\textbf{Mass Function}
The MF, computed using the $r-J$ colors of nearby stars \citep{golimowski08} and empirical mass-$M_J$ relations \cite{2000A&A...364..217D}, is shown in Figure \ref{mf_plot}.   The data points are well fit by a log--normal distribution with a characteristic mass of M $= 0.27 \pm 0.01 \msun$.

\begin{theacknowledgments}
  We gratefully acknowledges the support of NSF Grant AST 06-07644.
\end{theacknowledgments}





\IfFileExists{\jobname.bbl}{}
 {\typeout{}
  \typeout{******************************************}
  \typeout{** Please run "bibtex \jobname" to optain}
  \typeout{** the bibliography and then re-run LaTeX}
  \typeout{** twice to fix the references!}
  \typeout{******************************************}
  \typeout{}
 }

\end{document}